\documentclass{pasj01}
\Received{2016 May 07}
\Accepted{2016 July 08}
\Published{publication date}
\usepackage{natbib,color}

\begin{document}

\title{SXDF-ALMA 2 Arcmin$^2$ Deep Survey:\\ Resolving and Characterizing the Infrared Extragalactic Background Light Down to 0.5 mJy}
\author{Yuki \textsc{Yamaguchi}\altaffilmark{1}, Yoichi \textsc{Tamura}\altaffilmark{1}, Kotaro \textsc{Kohno}\altaffilmark{1,2}, Itziar \textsc{Aretxaga}\altaffilmark{3}, James S.~\textsc{Dunlop}\altaffilmark{4}, Bunyo \textsc{Hatsukade}\altaffilmark{5}, David \textsc{Hughes}\altaffilmark{3}, Soh \textsc{Ikarashi}\altaffilmark{6}, Shun \textsc{Ishii}\altaffilmark{1}, Rob J.~\textsc{Ivison}\altaffilmark{4,7}, Takuma \textsc{Izumi}\altaffilmark{1}, Ryohei \textsc{Kawabe}\altaffilmark{5,8,9}, Tadayuki \textsc{Kodama}\altaffilmark{5,9}, Minju \textsc{Lee}\altaffilmark{5,8}, Ryu \textsc{Makiya}\altaffilmark{10,11}, Yuichi \textsc{Matsuda}\altaffilmark{5,9}, Kouichiro \textsc{Nakanishi}\altaffilmark{5,9}, Kouji \textsc{Ohta}\altaffilmark{12}, Wiphu \textsc{Rujopakarn}\altaffilmark{10,13}, Ken-ichi \textsc{Tadaki}\altaffilmark{14}, Hideki \textsc{Umehata}\altaffilmark{1}, Wei-Hao \textsc{Wang}\altaffilmark{15}, Grant W.~\textsc{Wilson}\altaffilmark{16}, Kiyoto \textsc{Yabe}\altaffilmark{10}, and Min S.~\textsc{Yun}\altaffilmark{16}} %
\altaffiltext{1}{%
Institute of Astronomy, The University of Tokyo, 2-21-1, Osawa, Mitaka, Tokyo 181-0015}
\altaffiltext{2}{Research Center for the Early Universe, The University of Tokyo, 7-3-1, Hongo, Bunkyo, Tokyo 113-0033}
\altaffiltext{3}{Instituto Nacional de Astrof\'{i}sica, \'{O}ptica y Electr\'{o}nica, Tonantzintla, Luis Enrique Erro 1, Sta. Ma. Tonantzintla, Puebla, M\'{e}xico}
\altaffiltext{4}{Institute for Astronomy, University of Edinburgh, Royal Observatory, Blackford Hill, Edinburgh, EH9 3HJ, UK}
\altaffiltext{5}{National Astronomical Observatory of Japan, 2-21-1 Osawa, Mitaka, Tokyo 181-8588}
\altaffiltext{6}{Kapteyn Astronomical Institute, University of Groningen, P.O. Box 800, 9700AV Groningen, The Netherlands}
\altaffiltext{7}{European Southern Observatory, Karl-Schwarzschild-Str. 2, D-85748 Garching, Germany}
\altaffiltext{8}{Department of Astronomy, School of Science, University of Tokyo, Bunkyo, Tokyo 113-0033}
\altaffiltext{9}{The Graduate University for Advanced Studies (SOKENDAI), 2-21-1 Osawa, Mitaka, Tokyo 181-8588}
\altaffiltext{10}{Kavli Institute for the Physics and Mathematics of the universe, Todai Institutes for Advanced Study, the University of Tokyo, Kashiwa, 277-8583 Japan (Kavli IPMU, WPI)}
\altaffiltext{11}{Max-Planck-Institut für Astrophysik, Karl-Schwarzschild Str. 1, D-85741 Garching, Germany}
\altaffiltext{12}{Department of Astronomy, Kyoto University, Sakyo, Kyoto 606-8502}
\altaffiltext{13}{Department of Physics, Faculty of Science, Chulalongkorn University, 254 Phayathai Road, Pathumwan, Bangkok 10330, Thailand}
\altaffiltext{14}{Max-Planck-Institut f\"{u}r Extraterrestrische Physik (MPE), Giessenbachstr., D-85748 Garching, Germany}
\altaffiltext{15}{Institute of Astronomy and Astrophysics, Academia Sinica, Taipei 10617, Taiwan}
\altaffiltext{16}{Department of Astronomy, University of Massachusetts, Amherst, MA 01003, USA}
\email{yyamaguchi@ioa.s.u-tokyo.ac.jp}

\maketitle

\begin{abstract}
We present a multi-wavelength analysis of five submillimeter sources ($S_\mathrm{1.1mm}$ = 0.54--2.02 mJy) that were detected during our 1.1-mm-deep continuum survey in the SXDF-UDS-CANDELS field (2 arcmin$^2$, 1$\sigma$ = 0.055 mJy beam$^{-1}$) using the Atacama Large Millimeter/submillimeter Array (ALMA). The two brightest sources correspond to a known single-dish (AzTEC) selected bright submillimeter galaxy (SMG), whereas the remaining three are faint SMGs newly uncovered by ALMA. If we exclude the two brightest sources, the contribution of the ALMA-detected faint SMGs to the infrared extragalactic background light is estimated to be $\sim 4.1^{+5.4}_{-3.0}$ Jy deg$^{-2}$, which corresponds to $\sim 16^{+22}_{-12}\%$ of the infrared extragalactic background light. This suggests that their contribution to the infrared extragalactic background light is as large as that of bright SMGs. We identified multi-wavelength counterparts of the five ALMA sources. One of the sources (SXDF-ALMA3) is extremely faint in the optical to near-infrared region despite its infrared luminosity ($L_\mathrm{IR}\simeq 1\times10^{12} \LO$ or SFR $\simeq$ 100 $\MO$ yr$^{-1}$). By fitting the spectral energy distributions (SEDs) at the optical-to-near-infrared wavelengths of the remaining four ALMA sources, we obtained the photometric redshifts ($z_\mathrm{photo}$) and stellar masses ($M_*$): $z_\mathrm{photo} \simeq 1.3$--$2.5$, $M_* \simeq (3.5$--$9.5) \times10^{10}\ \MO$. We also derived their star formation rates (SFRs) and specific SFRs (sSFRs) as $\simeq 30$--$200 \MO$ yr$^{-1}$ and $\simeq 0.8$--$2$ Gyr$^{-1}$, respectively. These values imply that they are main-sequence star-forming galaxies.
\end{abstract}

\section{Introduction}
Determining the contributors of dust-obscured galaxies to the cosmic star-formation rate density (cosmic SFRD) is a major goal of deep surveys at far-infrared, millimeter, submillimeter, and radio wavelengths. In fact, deep surveys using the {\it Infrared Space Observatory} ({\it ISO}), {\it AKARI}, and the {\it Herschel Space Observatory} ({\it Herschel}) have revealed that dusty star-forming galaxies largely dominate the cosmic SFRD up to the redshift $z\approx$ 1--3 \citep[e.g.,][]{takeuchi2005, goto2011, brugarella2013}.

Over the past decade, a series of wide-area surveys performed at millimeter/submillimeter wavelengths using single-dish telescopes have revealed many bright submillimeter galaxies (SMGs) \citep[e.g.,][and references therein]{samil1997, hughes1998, barger1998, blain2002, greve2004, weiss2009, scott2010, hatsukade2011, casey2013, umehata2014} with observed flux densities larger than a few mJy at millimeter/submillimeter wavelengths. They have large total infrared (IR; {rest-frame 8--1000 $\micron$}) luminosities ($L_{\rm IR}\sim10^{12-13}\ \LO$) powered by dust-obscured star formation \citep[e.g.,][]{alexamder2005, laird2010}, and their redshift distribution peaks are at $z\approx$ 2.2--2.5 \citep[e.g.,][]{capman2005,simpson2014}. Their extreme star formation rates (SFRs $\gtrsim$ a few 100--1000 $\MO$yr$^{-1}$) make them non-negligible contributors to cosmic star formation \citep[e.g.,][]{hughes1998, casey2013, wardlow2011, swinbank2014}. However, the contribution of SMGs detected by single-dish surveys to the infrared extragalactic background light, which is believed to be the integrated infrared emissions from all extragalactic sources along the line of sight, is 20--40\% at 850 $\micron$ \citep[e.g.,][]{eales1999, coppin2006, weiss2009} and 10--20\% at 1.1 mm \citep[e.g.,][]{hatsukade2011, scott2012}. Thus, the bulk of infrared extragalactic background light remains unresolved with single-dish telescopes.

By using stacking analysis of $K$-selected galaxies, \citet{greve2010} found that these galaxies contribute $\simeq16.5\%$ to the infrared extragalactic background light at 870 $\micron$, although individual source properties remained unexplored in this stacking analysis.

The advent of the Atacama Large Millimeter/submillimeter Array (ALMA), which offers high sensitivity and angular resolution capabilities, has allowed a fainter population of SMGs to be unveiled below the confusion limit of single-dish telescopes. Here, we refer to the fainter population of SMGs with flux densities of $\sim$ 0.1--1 mJy at 1.1--1.3 mm as ``faint SMGs.'' Their estimated contributions to the infrared extragalactic background light are $\simeq$ 50\%--80\% \citep{hatsukade2013, ono2014, oteo2015}. Deeper number counts down to $\sim$ 0.02 mJy have recently obtained by \citet{carniani2015} and \citet{fujimoto2015}, who claimed that $\sim$ 100\% of the infrared extragalactic background light is resolved at 1.2--1.3 mm. These results suggest that faint SMGs can play an important role in the cosmic star-formation activities at high redshifts. However, their contributions to the cosmic SFRD are still unknown because of the lack of redshift information.

Single-dish telescopes have been used in attempts to detect faint SMGs with the aid of gravitational magnification by lensing clusters. For example, \citet{knudsen2008} constrained the faint end ($S_\mathrm{850 \micron} \simeq 0.1$ mJy) of the 850 $\micron$ number counts by using cluster magnification. \citet{chen2014} performed follow-up observations of these lensed faint SMGs by using the submillimeter array \citep[SMA;][]{moran1998} and implied that there are many faint SMGs that are faint at optical/near-infrared wavelengths and have been missed in deep optical/near-infrared surveys. In lens surveys, however, the effective sensitivity comes at the cost of a reduced survey volume; the effective (source-plane) area within sufficient magnification for faint SMG detection is only $\sim 0.1$ arcmin$^2$ for a typical rich cluster \citep{knudsen2008}.\footnote{\cite{fujimoto2015} also used gravitational magnification from lensing clusters, and their survey area was $\sim$ 0.5 arcmin$^2$ (source-plane).} This also increases the cosmic variance uncertainty \citep[e.g.,][]{robertson2014}. Therefore, it is still necessary to obtain wide ($> 1$ arcmin$^2$) and deep (1$\sigma\lesssim$ 0.1 mJy) blank/unlensed field surveys at a higher angular resolution to gain a better understanding of faint SMGs and their true contributions to the cosmic SFRD.

Another key issue to understand galaxy evolution is the star formation properties of galaxies. Star-forming galaxies have a correlation between their stellar masses and SFRs, which is defined as a main sequence \citep[e.g.,][]{daddi2007,rodighiero2011,rodighiero2014,schriber2015}. Main sequence star-forming galaxies are ``normal'' star-forming galaxies selected by optical/near-infrared colors [e.g., {\it BzK} galaxies, typical specific SFRs (sSFRs) $\sim1$ Gyr$^{-1}$ at $z\sim$ 1.4--2.5; \citet{rodighiero2011}]. However, outliers of the correlation exist with higher sSFRs than those of main sequence star formation galaxies \citep[sSFRs $\gtrsim10^{1}$--$10^2$ Gyr$^{-1}$ at $z\sim$ 1.4--2.5;][]{rodighiero2011}. These outliers are often referred to as starburst galaxies. Many bright SMGs are classified as starburst galaxies or the high-mass end of main sequence galaxies \citep[e.g.,][]{takagi2008, dacunha2015}. However, it is not understood whether faint SMGs are on or above the main sequence because the stellar masses of these faint SMGs have not yet been measured. Understanding the star-forming properties of faint SMGs is also helpful to unveil the evolution of the cosmic SFRD because they are thought to be the main contributor to the infrared extragalactic background light \citep[e.g.,][]{carniani2015,fujimoto2015}.

In this paper, we present spectral energy distributions (SEDs) for optical-to-radio counterparts to five submillimeter sources that were detected in our 2 arcmin$^2$ 1.1-mm-deep survey of the Subaru/\textit{XMM-Newton} Deep Survey Field \citep[SXDF;][]{furusawa2008} using ALMA (Project ID: 2012.1.00756.S, PI: K. Kohno) to understand their contribution to the infrared extragalactic background light and cosmic SFRD. We also discuss their multi-wavelength properties. Our survey field was also covered by the UKIRT Infrared Deep Sky Survey Ultra-Deep Survey \citep[UDS;][]{lawrence2007} and Cosmic Assembly Near-infrared Deep Extragalactic Legacy Survey \citep[CANDELS;][]{grogin2011,Koekemoer2011}.

This paper is structured as follows. Section \ref{multi-image} presents the ALMA observations and multi-wavelength data used in this study. Section \ref{counterpart} presents the results of the multi-wavelength counterpart identification of the ALMA sources. Section \ref{SED} derives the photometric redshift, stellar mass ($M_*$), SFRs, and sSFRs and presents the optical-to-radio SEDs. Sections \ref{discussion} and \ref{summary} are devoted to the discussion and summary, respectively. Throughout this paper, we assume a $\Lambda$ cold dark matter cosmology with $\Omega_M = 0.3$, $\Omega_\Lambda = 0.7$, and $H_0 =$ 70 km s$^{-1}$ Mpc$^{-1}$. All magnitudes are given according to the AB system.

\section{Multi-wavelength images}
\label{multi-image}
\subsection{ALMA observations and source identifications}
\label{ALMA}
Here, we briefly summarize the ALMA data; the details will be given in a subsequent paper \citep[Kohno et al.\ in preparation; see also][]{tadaki2015, kohno2016, hatsukade2016}. The ALMA observations were carried out on July 17 and 18, 2014 (Cycle 1). The 2 arcmin$^2$ map of the SXDF-UDS-CANDELS field was obtained at 274 GHz (or 1.1 mm; Band 6). The region is centered at $(\alpha,\,\delta)_\mathrm{J2000} =$ (\timeform{02h17m41.31s}, \timeform{-05D13'28''.9}) and is covered by 19 pointings of ALMA. For the observations, 30--32 antennas were employed, and the array was close to the C32-4 configuration, which has minimum and maximum baselines of 20 and 650 m, respectively (the synthesized beam is 0.53 arcsec $\times$ 0.41 arcsec with a position angle of \timeform{64D} in the naturally weighted map). We performed our observations under good conditions, where the precipitable water vapor was in the range of 0.42--0.55 mm. The phase and bandpass were calibrated with J0215-0222 and J0241-0815, respectively. The flux was calibrated with J2258-279 and J0238+166. The absolute calibration accuracy for Cycle 1 is 10\% for Band 6 (ALMA Cycle 1 Technical Handbook). The data were processed with the Common Astronomy Software Application \citep[CASA;][]{mcmullin2007}. The map was processed with the \texttt{CLEAN} algorithm \citep{hogdom1974} using natural weighting.

The resulting image had a $1\sigma$ sensitivity of 0.048--0.061 mJy beam$^{-1}$, and the typical noise level was 0.055 mJy beam$^{-1}$ \citep[][Kohno et al. in preparation]{hatsukade2016}. From the ALMA map, we extracted five significant sources with a signal-to-noise ratio of S/N $> 5$ \citep[hereafter SXDF-ALMA1, 2, 3, 4, and 5; table \ref{table1}, see also][]{kohno2016}\footnote{\cite{hatsukade2016} report an additional source with S/N = 5, which is detected as SXDF-ALMA6 with S/N = 4.7 in \cite{kohno2016}. We do not discuss this source in multi-wavelength context here, because it has no counterpart at multi-wavelength images.}. Two of the five sources (SXDF-ALMA1 and 2) were detected as a single bright SMG ($S_\mathrm{1.1\ mm}$ = 3.5$^{+0.6}_{-0.5}$ mJy) with the Astronomical Thermal Emission Camera \citep[AzTEC;][]{wilson2008}/Atacama Submillimeter Telescope Experiment \citep[ASTE;][]{ezawa2004} 1.1 mm survey in SXDF (Ikarashi et al.\ in preparation).

\begin{figure*}[t]
\begin{center}
\FigureFile(160mm,50mm){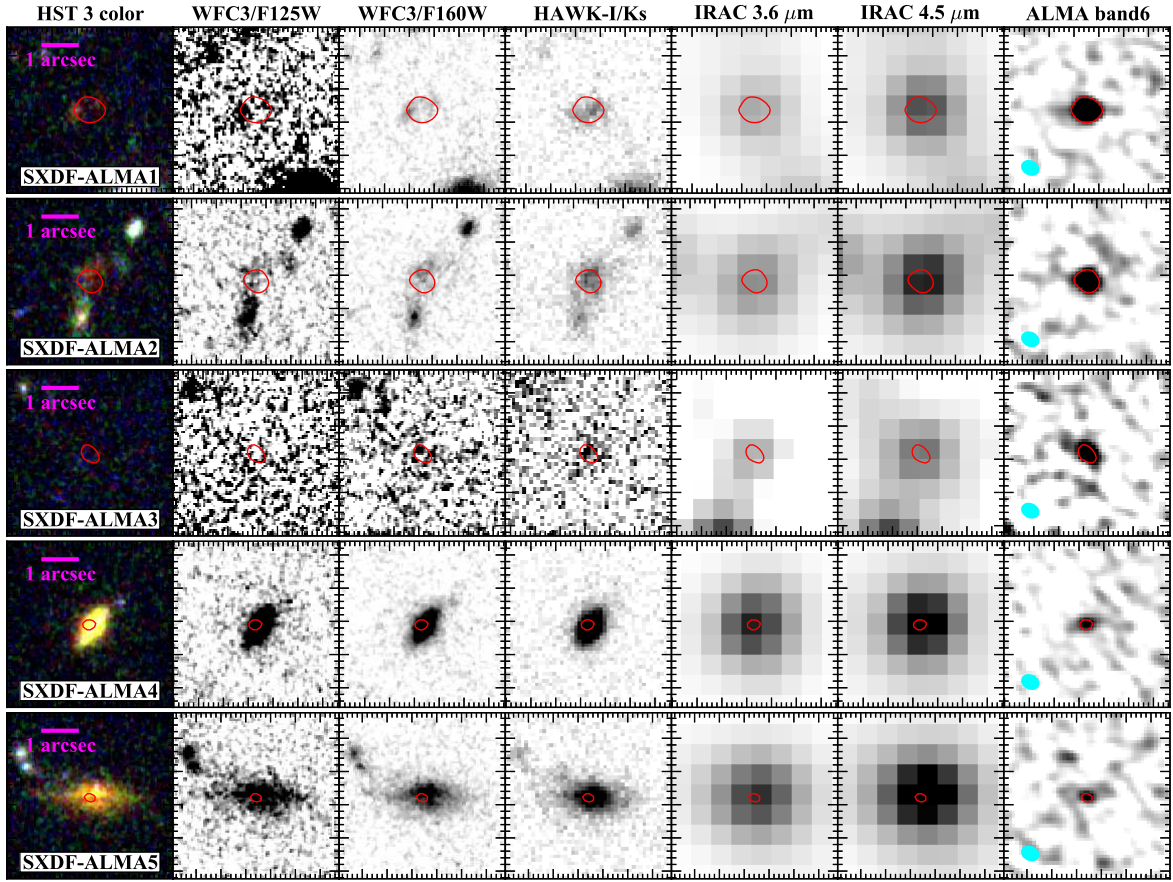}
\end{center}
\caption{Optical-to-near-infrared images of our ALMA sources. From left to right: {\it HST} 3 color (R: WFC3/{\it F160W}, G: WFC3/{\it F125W}, B: ACS/{\it F814W}), {\it HST} WFC3/{\it F125W}, {\it HST} WFC3/{\it F160W}, VLT HAWK-I/{\it Ks}, {\it Spitzer} IRAC/3.6 $\micron$, {\it Spitzer} IRAC/4.5 $\micron$, and ALMA 1.1 mm images (5 arcsec $\times$ 5 arcsec). The solid red contours indicate the ALMA 1.1 mm detection at the 5$\sigma$ level. The synthesized beams are presented in the bottom left of the ALMA images (cyan).}\label{stamp}
\end{figure*}

\subsection{Optical-to-near-infrared images}
In order to characterize the stellar properties of the ALMA sources, we used archival optical-to-near-infrared images collected by ground-based and space-borne facilities such as Subaru, the Very Large Telescope (VLT), the {\it Hubble Space Telescope} ({\it HST}), and the {\it Spitzer} Space Telescope ({\it Spitzer}). We describe the details of the optical/near-infrared data below.

\subsubsection{Ground-based telescopes}
Optical ground-based imaging observations of the UDS field were made with Subaru/Suprime-Cam \citep{miyazaki2002} using five wideband filters (BVRci'z') as part of Subaru/{\it XMM-Newton} Deep Survey \citep{furusawa2008}. These data reached 3$\sigma$ limiting magnitudes of $B = 28.4$, $V = 27.8$, $Rc = 27.7$, $i' = 27.7$, and $z' = 26.6$ with a 1 arcsec radius aperture \citep{furusawa2008}.

The CANDELS UDS field was also observed as part of the HAWK-I UDS and GOODS-S survey \citep[HUGS; VLT large program ID: 186.A-0898,][]{fontana2014} with two near-infrared broadband filters ($Y$ and $Ks$). The data reached 5$\sigma$ limiting magnitudes of $Y = 27.05$ and $Ks = 26.16$ with 0.42 and 0.36 arcsec radius apertures \citep{galametz2013}.

\subsubsection{{\it HST}}
The {\it HST} data were taken with the Advanced Camera for Surveys \citep[ACS;][]{ford1998}/{\it F606W} and ACS/{\it F814W} together with the Wide Field Camera 3 \citep[WFC3;][]{kimble2008}/{\it F125W} and WFC3/{\it F160W}. The final UDS {\it HST} images are publicly available via the STScI archive.\footnote{https://archive.stsci.edu/prepds/candels/} These images reached 5$\sigma$ limiting magnitudes of ACS/{\it F606W} = 26.74, ACS/{\it F814W} = 26.67, WFC3/{\it F125W} = 26.80, and WFC3/{\it F160W} = 26.91 with a 0.7 arcsec radius aperture, respectively \citep{Koekemoer2011}.

\subsubsection{{\it Spitzer}/IRAC}
We use {\it Spitzer}/InfraRed Array Camera \citep[IRAC;][]{fazio2004} data. Channel 1 (3.6 $\micron$) and Channel 2 (4.5 $\micron$) data were from the {\it Spitzer} Extended Deep Survey \citep[SEDS; PI: G. Fazio;][]{ashby2013}, and Channel 3 (5.6 $\micron$) and Channel 4 (8.0 $\micron$) data were from the {\it Spitzer} UKIDSS Ultra Deep Survey \citep[SpUDS; PI: J. Dunlop;][]{caputi2011}. These images reached 5$\sigma$ limiting magnitudes of 24.72, 24.61, 22.30, and 22.26 with 1.9, 1.9, 2.08, and 2.20 arcsec radius apertures at 3.6, 4.5, 5.8, and 8.0 $\micron$, respectively \citep{galametz2013}.

\subsection{mid-infrared-to-radio images}
Here, we summarize mid-infrared to radio images, which we retrieved from public archives. 

\subsubsection{{\it Spitzer}/MIPS}
We also used the Multiband Imaging Photometer for the {\it Spitzer} \citep[MIPS;][]{rieke2004} 24 $\micron$ image from SpUDS. The final {\it Spitzer}/MIPS image is publicly available via the NASA/IPAC Infrared Science Archive.\footnote{http://irsa.ipac.caltech.edu/data/SPITZER/SpUDS/} The image reached a 3$\sigma$ limiting magnitude of 19.9 (details of our photometry are explained in section \ref{photo_mir-r}).

\subsubsection{{\it Herschel}/PACS and SPIRE}
The far-infrared to submillimeter {\it Herschel} images were taken with the Photodetector Array Camera and Spectrometer (PACS) at 100 and 160 $\micron$ \citep{poglitsch2010} and with the Spectral and Photometric Imaging Receiver (SPIRE) at 250, 350, and 500 $\micron$ \citep{griffin2010} as part of the {\it Herschel} Multi-tiered Extragalactic Survey \citep[HerMES\footnote{http://hermes.sussex.ac.uk/}; see][for details]{oliver2012}. These images were retrieved from the {\it Herschel} Science Archive. The 5$\sigma$ instrument sensitivity (ignoring confusion noise) was 6.8, 12.9, 11.2, 9.3, and 13.4 mJy at 100, 160, 250, 350, and 500 $\micron$, respectively \citep{oliver2012}. Note that {\it Herschel}/SPIRE on sky images is dominated by confusion noise because of the large beam size \citep[18.1 arcsec, 25.2 arcsec, and 36.6 arcsec for 250, 350, and 500 $\micron$;][]{griffin2010}. In HerMES, the confusion noise of {\it Herschel}/SPIRE is calculated from images of the GOODS-N, Lockman-North, and Lockman-SWIRE fields \citep{nguyen2010}. Here, we adopted their confusion noise of $5\sigma$ = 24.0, 27.5, and 30.5 mJy at 250, 350, and 500 $\micron$, respectively.

\subsubsection{Radio images}
Radio images at 1.4 GHz and 6 GHz were obtained by using the Very Large Array (VLA) and the Karl G.~Jansky Very Large Array (JVLA), respectively.

The details on the VLA 1.4 GHz observations are given by Arumugam et al.~(in preparation). The synthesized beam size for this 1.4 GHz image was 1.8 arcsec $\times$ 1.6 arcsec with a position angle of the \timeform{-3D}. The 1.4 GHz image reached a 1$\sigma$ uncertainty of 8 $\mu$Jy. On the other hand, the JVLA 6 GHz observations can be presented by Tadaki et al.~(in preparation). The synthesized beam size was 0.5 arcsec $\times$ 0.4 arcsec with a position angle of \timeform{-3D}. The 6 GHz image reached a 1$\sigma$ uncertainty of 0.72 $\mu$Jy.

\subsection{X-ray images}
\cite{ueda2008} presented an X-ray source catalog for the SXDF-UDS-CANDELS field using the \textit{XMM-Newton} satellite. The sensitivity limit of the catalog reached 6 $\times$ 10$^{-16}$, 8 $\times$ 10$^{-16}$, 3 $\times$ 10$^{-15}$, and 5 $\times$ 10$^{-15}$ erg cm$^{-2}$ s$^{-1}$ in the 0.5--2, 0.5--4.5, 2--10, and 4.5--10 keV bands, respectively. The five ALMA sources are not listed in the catalog. 

\section{Counterpart identification and photometry}
\label{counterpart}
\subsection{Optical-to-near-infrared counterparts and photometry}
\label{photo_opt_nir}
Figure \ref{stamp} shows the {\it HST} 3 color image (for red: WFC3/{\it F160W}, for green: WFC3/{\it F125W}, for blue: ACS/{\it F814W}), WFC3/{\it F125W}, WFC3/{\it F160W}, HAWK-I/{\it Ks}, {\it Spitzer}/IRAC 3.6 $\micron$, 4.5 $\micron$, and ALMA Band 6 images overlaid with the ALMA contours. All five ALMA sources had counterparts in at least four independent bands. The WFC3/{\it F160W} counterparts to four of the five ALMA sources (SXDF-ALMA1, 2, 4, and 5) were already identified in the catalog presented by \citet{galametz2013} ($H \simeq$ 22.4--24.5). Although SXDF-ALMA3 is not cataloged in \citet{galametz2013}, it was marginally detected (4.2$\sigma$, $H = 25.30 \pm 0.25$) with WFC3/{\it F160W} (details of our photometry are explained later in this section). We found that SXDF-ALMA1 and SXDF-ALMA2 coincide with the cataloged H$\alpha$ emitters (HAEs) having narrowband redshifts of $z=2.53\pm0.02$ \citep{tadaki2013}. 

We used the Image Reduction and Analysis Facility \cite[\texttt{IRAF;}][]{tody1993} to measure the flux densities of the counterparts in the optical-to-near-infrared images. First, to account for the point spread function (PSF) difference between images, images were PSF-matched using the \texttt{IRAF} task \texttt{GAUSS}. Except for the {\it Spitzer}/IRAC images, we adopted the Gaussian PSF with a full width at half maximum of 1 arcsec. {\it Spitzer}/IRAC images were PSF-matched to the 8.0 $\micron$ band image, which had the poorest angular resolution among the IRAC bands at 2.2 arcsec.

Next, we performed optical-to-near-infrared photometry with a 2 arcsec diameter aperture at the position of the ALMA sources using the \texttt{IRAF} task \texttt{APPHOT}. We corrected the aperture correction using the same procedure described by \citet{ono2010}. 
We measured fluxes for 20 bright point sources in a series of diameter apertures from 2 arcsec up to 6 arcsec with an interval of 0.1 arcsec. Since we found that the fluxes level off for $>$ 5 arcsec diameter apertures, we defined 5 arcsec diameter aperture magnitudes as total magnitudes. Then, we selected 100 point sources, measured fluxes over 2 and 5 arcsec diameter apertures, and calculated an accurate offset between these two aperture magnitudes as the aperture correction term. We estimated the flux uncertainties by using the \texttt{SDFRED} \citep{yagi2002, ouchi2004} task \texttt{LIMITMAG}. Table \ref{table1} lists the photometry results.

Note that \citet{tadaki2015} used the optical-to-near-infrared photometry of SXDF-ALMA1 and 2 from the photometry catalog presented by \citet{skelton2014}, which follows a different photometric procedure from ours. We did not use their photometry because we preferred comparing all of the ALMA sources in the same manner.

\subsection{mid-infrared-to-radio counterparts and photometry}
\label{photo_mir-r}
\begin{figure*}[t]
\begin{center}
\FigureFile(160mm,50mm){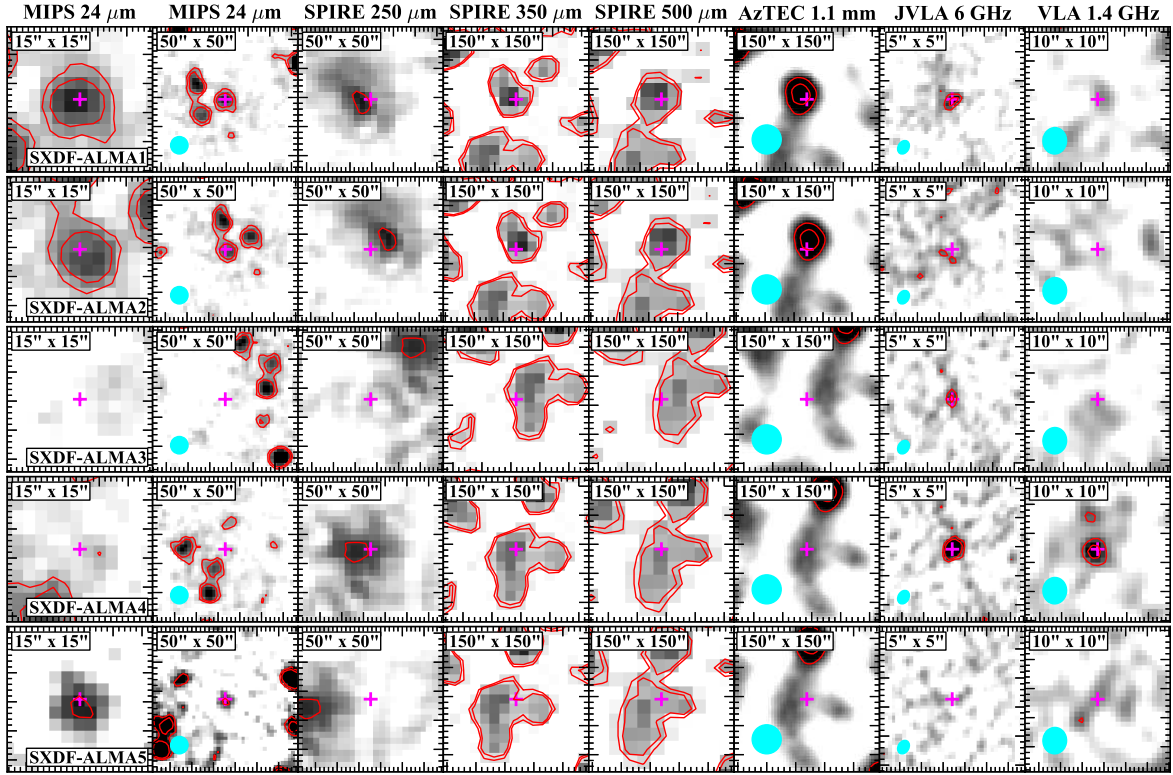}
\end{center}
\caption{mid-infrared-to-radio images of SXDF-ALMA1, 2, 3, 4, and 5 (from top to bottom). From left to right: 15 arcsec $\times$ 15 arcsec image of {\it Spitzer}/MIPS 24 $\micron$, 50 arcsec $\times$ 50 arcsec image of {\it Spitzer}/MIPS 24 $\micron$, {\it Herschel}/SPIRE 250 $\micron$, 150 arcsec $\times$ 150 arcsec image of {\it Herschel}/SPIRE 350, 500 $\micron$, AzTEC/ASTE 1.1 mm, 5 arcsec $\times$ 5 arcsec image of JVLA 6 GHz, and 10 arcsec $\times$ 10 arcsec image of VLA 1.4 GHz, respectively. The magenta crosses mark the ALMA positions. The beam sizes of MIPS 24 $\micron$, AzTEC/ASTE, JVLA, and VLA are shown by cyan symbols. Red solid lines indicate the contours (3$\sigma$ and 5 $\sigma$) of images.}
\label{multi_ww}
\end{figure*}

Figure \ref{multi_ww} shows {\it Spitzer}/MIPS 24 $\micron$, {\it Herschel}/SPIRE 250, 350, 500 $\micron$, AzTEC/ASTE 1.1 mm, JVLA 6 GHz, and VLA 1.4 GHz images of the ALMA sources. For the MIPS 24 $\micron$ image, we performed aperture photometry with a 3 arcsec aperture radius. The MIPS instrument handbook\footnote{http://irsa.ipac.caltech.edu/data/SPITZER/docs/mips/mipsinstrumenthandbook/} was used to calculate the aperture correction for the missing flux outside the aperture. For the {\it Herschel}/SPIRE bands, we used the {\it Herschel} Interactive Processing Environment \citep[HIPE;][]{ott2010} to perform the \texttt{sourceExtractorSussextractor} task at the ALMA source positions. Note that, because the SPIRE photometry listed in table \ref{table1} is highly likely to be affected by blending with nearby infrared sources, the nominal values should be considered as upper limits. In the VLA 1.4 GHz and JVLA 6 GHz images we measured the flux densities with two-dimensional Gaussian fitting by using the Astronomical Image Processing System \citep[\texttt{AIPS;}][]{van1996} to perform the \texttt{JMFIT} task. For non-detection, we set a 3$\sigma$ upper limits. Table \ref{table1} summarizes the results. We present the mid-infrared-to-radio properties of individual ALMA sources below.
\bigskip

\textit{SXDF-ALMA1 and 2}: SXDF-ALMA1 and 2 are close to each other; the separation is $\sim 10$ arcsec, which may cause significant blending with each other in the mid-infrared-to-far-infrared bands. They were both detected at 24 $\micron$ ($18.66\pm0.12$ mag and $18.82\pm0.13$ mag, respectively). In the SPIRE images, it is likely that SXDF-ALMA1, 2, and a nearby MIPS source are blended, which resulted in a single cataloged SPIRE source of J021740.9$-$051309 \citep{oliver2012}. Therefore, we simply placed $5\sigma$ upper limits on the SPIRE photometry at the positions of SXDF-ALMA1 and 2. The sum of the 1.1 mm flux density of the two sources ($S_\mathrm{1.1mm} = 3.4 \pm 0.2$ mJy) showed good agreement with the flux density of the blended AzTEC/ASTE source ($S_\mathrm{1.1mm} = 3.5^{+0.6}_{-0.5}$ mJy; Ikarashi et al.~in preparation). SXDF-ALMA1 was detected at 6 GHz ($S_\mathrm{6 GHz}=12.0\pm2.9$ $\mu$Jy), but SXDF-ALMA2 had no counterpart in the radio images.

\textit{SXDF-ALMA3:} SXDF-ALMA3 was only detected at 6 GHz ($S_\mathrm{6 GHz}=10.0\pm2.8$ $\mu$Jy). The interpretation is given in the following sections.

\textit{SXDF-ALMA4:} This object was detected not at 24 $\micron$ but at 6 GHz and 1.4 GHz with $23.0\pm2.5$ $\mu$Jy and $50.09\pm7.45$ $\mu$Jy, respectively. In SPIRE bands, this source was heavily blended with a nearby 24 $\micron$ source (with a separation of $\sim 8$ arcsec), and the blended source was identified as J021742.5$-$051406 \citep{oliver2012}.

\textit{SXDF-ALMA5:} SXDF-ALMA5 was detected in the MIPS 24 $\micron$ band ($19.53\pm0.26$ mag), although it was not detected in the {\it Herschel} and radio images.

\begin{figure*}[!ht]
\begin{center}
\FigureFile(160mm,50mm){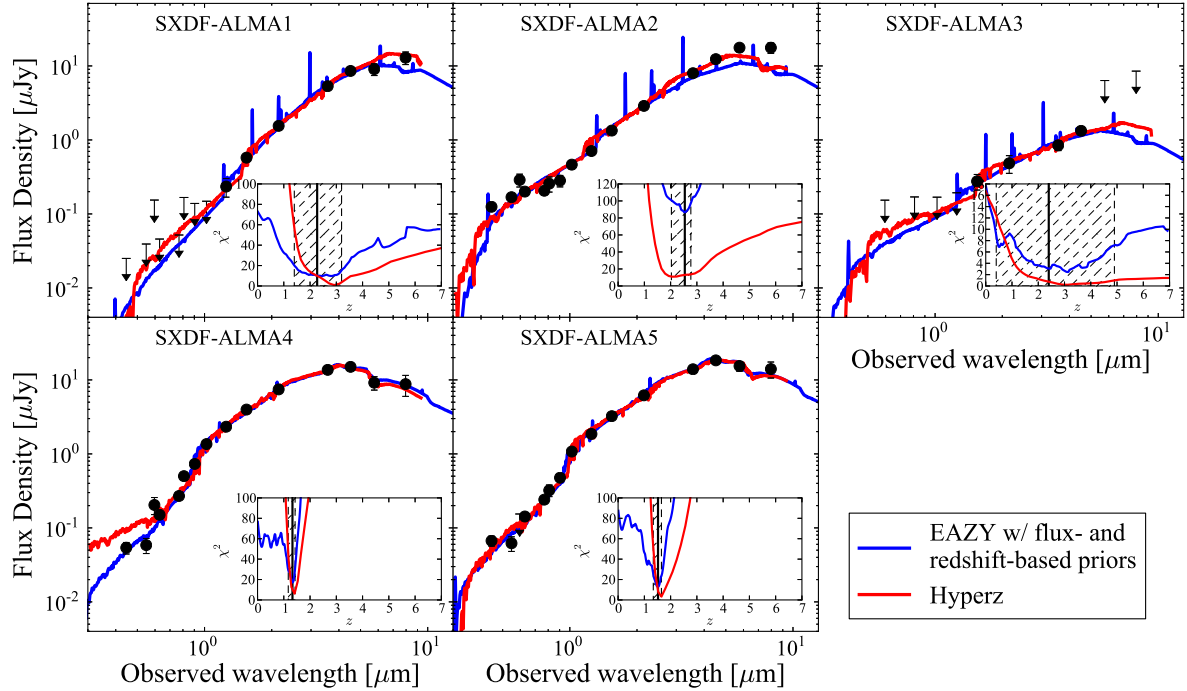}
\end{center}
\caption{Photometry and best-fit SEDs for SXDF-ALMA1, 2, 3, 4, and 5. The dots with error bars are photometric data points, whereas the arrows indicate 3$\sigma$ upper limits. The blue and red lines indicate the results calculated with \texttt{EAZY} and \texttt{HYPERZ}, respectively. The inset panels in each plot show the $\chi^2$ distribution as a function of the redshift and indicate the best-fit photometric redshift as estimated by \texttt{EAZY} with a black solid line. The hatched region shows 99\% confidence intervals estimated with \texttt{EAZY}.}\label{sed}
\end{figure*}

\renewcommand{\arraystretch}{0.8}
\begin{table*}
\caption{Multi-wavelength photometry of our ALMA sources and their derived properties.}
\begin{tabular}{rlccccc}
\hline\hline
&&SXDF-ALMA1&SXDF-ALMA2&SXDF-ALMA3&SXDF-ALMA4&SXDF-ALMA5\\ \hline
R.A.$_{\mathrm{1.1 mm}}$\footnotemark[$*$] &(J2000)&\timeform{02h17m40s.52}&\timeform{02h17m41s.12}&\timeform{02h17m43s.64}&\timeform{02h17m42s.33}&\timeform{02h17m41s.23}\\
Dec.$_{\mathrm{1.1 mm}}$\footnotemark[$*$] &(J2000)&\timeform{-05D13'10''.64}&\timeform{-05D13'15''.19}&\timeform{-05D14'23''.81}&\timeform{-05D14'05''.09}&\timeform{-05D14'02''.73}\\
$S_{\mathrm{1.1 mm;\ peak}}$\footnotemark[$\dagger$]  &[mJy/beam]&1.69$\pm$0.06&0.92$\pm$0.07&0.84$\pm$0.09&0.36$\pm$0.05&0.28$\pm$0.05\\
$S_{\mathrm{1.1 mm;\ total}}$\footnotemark[$\ddagger$]  &[mJy]&2.02$\pm$0.12&1.38$\pm$0.14&1.16$\pm$0.19&0.54$\pm$0.12&0.56$\pm$0.14\\
R.A.$_{\mathrm{F160W}}$\footnotemark[$\S$]  &(J2000)&\timeform{02h17m40s.55}&\timeform{02h17m41s.12}&$-$&\timeform{02h17m42s.34}&\timeform{02h17m41s.22}\\
Dec.$_{\mathrm{F160W}}$\footnotemark[$\S$]  &(J2000)&\timeform{-05D13'10''.67}&\timeform{-05D13'14''.98}&$-$&\timeform{-05D14'05''.16}&\timeform{-05D14'02''.77}\\
Subaru/Suprime-Cam {\it B}  &[mag]&$>27.90$&$26.16\pm0.07$&$-$&$27.07\pm0.16$&$26.84\pm0.13$\\
Subaru/Suprime-Cam {\it V}  &[mag]&$>27.42$&$25.83\pm0.08$&$-$&$26.98\pm0.24$&$26.92\pm0.22$\\
Subaru/Suprime-Cam {\it Rc}  &[mag]&$>27.24$&$25.65\pm0.08$&$-$&$25.96\pm0.11$&$26.02\pm0.11$\\
Subaru/Suprime-Cam {\it i'}  &[mag]&$>27.11$&$25.62\pm0.09$&$-$&$25.32\pm0.07$&$25.45\pm0.08$\\
Subaru/Suprime-Cam {\it z'}  &[mag]&$>26.04$&$25.27\pm0.18$&$-$&$24.24\pm0.07$&$24.71\pm0.11$\\
{\it HST}/ACS {\it F606W}  &[mag]&$>25.93$&$25.25\pm0.19$&$>25.93$&$25.63\pm0.27$&$26.83\pm0.84$\\
{\it HST}/ACS {\it F814W}  &[mag]&$>25.84$&$25.38\pm0.24$&$>25.84$&$24.65\pm0.12$&$25.13\pm0.19$\\
{\it HST}/WFC3 {\it F125W}  &[mag]&$25.48\pm0.30$&$24.28\pm0.10$&$>25.66$&$22.98\pm0.03$&$23.23\pm0.04$\\
{\it HST}/WFC3 {\it F160W}  &[mag]&$24.50\pm0.12$&$23.58\pm0.05$&$25.30\pm0.25$&$22.40\pm0.02$&$22.63\pm0.02$\\
VLT/HAWK-I {\it Y}  &[mag]&$>25.97$&$24.74\pm0.12$&$>25.97$&$23.57\pm0.04$&$23.82\pm0.05$\\
VLT/HAWK-I {\it Ks}  &[mag]&$23.42\pm0.09$&$22.75\pm0.05$&$24.69\pm0.29$&$21.72\pm0.02$&$21.92\pm0.02$\\
{\it Spitzer}/IRAC 3.6 $\micron$  &[mag]&$22.08\pm0.02$&$21.65\pm0.02$&$24.08\pm0.14$&$21.06\pm0.01$&$21.04\pm0.01$\\
{\it Spitzer}/IRAC 4.5 $\micron$  &[mag]&$21.57\pm0.02$&$21.17\pm0.02$&$23.59\pm0.11$&$20.96\pm0.01$&$20.74\pm0.01$\\
{\it Spitzer}/IRAC 5.8 $\micron$  &[mag]&$21.51\pm0.17$&$20.78\pm0.24$&$>22.17$&$21.49\pm0.10$&$20.93\pm0.12$\\
{\it Spitzer}/IRAC 8.0 $\micron$  &[mag]&$21.12\pm0.16$&$20.78\pm0.12$&$>21.71$&$21.54\pm0.24$&$21.03\pm0.17$\\ 
{\it Spitzer}/MIPS 24 $\micron$ &[mag]&$18.66\pm0.12$&$18.82\pm0.13$&$>19.9$&$>19.9$&$19.53\pm0.26$\\
{\it Herschel}/PACS 100 $\micron$ &[mJy]&$<6.72$&$<6.72$&$<6.72$&$<6.72$&$<6.72$\\
{\it Herschel}/PACS 160 $\micron$ &[mJy]&$<12.8$&$<12.8$&$<12.8$&$<12.8$&$<12.8$\\
{\it Herschel}/SPIRE 250 $\micron$ &[mJy]&$(\mathit{17.9\pm5.0})$\footnotemark[$\|$]&$(\mathit{18.5\pm5.0})$\footnotemark[$\|$]&$<14.4$&$(\mathit{21.1\pm5.0})$\footnotemark[$\|$]&$<14.4$\\
{\it Herschel}/SPIRE 350 $\micron$ &[mJy]&$(\mathit{19.9\pm5.7})$\footnotemark[$\|$]&$(\mathit{21.4\pm5.7})$\footnotemark[$\|$]&$<16.5$&$(\mathit{19.3\pm5.7})$\footnotemark[$\|$]&$<16.5$\\
{\it Herschel}/SPIRE 500 $\micron$ &[mJy]&$(\mathit{15.3\pm6.3})$\footnotemark[$\|$]&$(\mathit{15.3\pm6.3})$\footnotemark[$\|$]&$<18.3$&$(\mathit{13.0\pm6.3})$\footnotemark[$\|$]&$<18.3$\\
JVLA 6 GHz &[$\mu$Jy]&$12.0\pm2.9$&$<2.16$&$10.9\pm2.8$&$23.0\pm2.5$&$<2.16$\\
VLA 1.4 GHz &[$\mu$Jy]&$<24$&$<24$&$<24$&$50.9\pm7.45$&$<24$\\
\hline
\multicolumn{6}{l}{\hbox to 0pt{\parbox{180mm}{{\bf Notes.}{\\ \footnotesize  footnotesize Inequality signs represent the 3$\sigma$ limits of photometry. We did not use Subaru data in SXDF-ALMA3 because of the contamination from a nearby source.
\\ \footnotemark[$*$] The sky positions of our ALMA sources adopted from Kohno et al.\ (in preparation).
\\ \footnotemark[$\dagger$] The 1.1 mm observed peak flux densities (primary beam collected) presented in Kohno et al.\ (in preparation).
\\ \footnotemark[$\ddagger$] The 1.1 mm observed spatially integrated flux densities (primary beam collected) presented in Kohno et al.\ (in preparation).
\\ \footnotemark[$\S$] The sky positions of {\it F160W} counterparts derived from {\it F160W}-selected catalog \citep{galametz2013}.
\\ \footnotemark[$\|$] SPIRE photometry using the HIPE task \texttt{sourceExtractorSussextractor} at the ALMA position. The errors were estimated by adding the 1$\sigma$ confusion errors and instrumental errors in the quadrature.
}}
\hss}}
\label{table1}
\end{tabular}
\end{table*}
\renewcommand{\arraystretch}{1.0}

\section{Multi-wavelength SED}
\label{SED}
\subsection{SED fitting at optical-to-near-infrared wavelengths}
\subsubsection{Photometric redshift calculation}
\label{photo_z}
We estimated the photometric redshifts of the ALMA sources.
To derive the photometric redshifts and check their reliability, we used two different SED fitting codes: \texttt{HYPERZ} \citep{bolzonella2000} and \texttt{EAZY} \citep{brammer2008}. They both compute the $\chi^2$ statistic for a set of SED models to the observed photometry, but \texttt{EAZY} includes the optional flux- and redshift-based priors \citep[see][for details]{brammer2008}. Confidence intervals were obtained by integrating the posterior redshift probability distributions \citep[see][for details]{brammer2008}. The following parameters were considered for SED fitting. The redshift range was set to $z=$ 0--7. Extinction is considered with the range of $A_{\rm V}=$ 0--5 mag in increments of 0.5 mag, and we adopted the Calzetti extinction law \citep{calzetti2000}. For \texttt{HYPERZ}, we utilized the SED templates of \citet{bc93} for elliptical, Sb, burst, constant, and star formation (Im). For \texttt{EAZY}, SED templates of \citet[][hereafter BC03]{bc03} were used. In the case of non-detection, we adopted the nominal photometric value with 1$\sigma$ uncertainty during the SED fitting.

Table \ref{table2} summarizes the derived photometric redshifts. Figure \ref{sed} shows the best-fit SEDs of the ALMA sources. Throughout this paper, we use the 99\% confidence intervals to represent the uncertainty of the photometric redshift estimates, as done in previous works \citep[e.g.,][]{wardlow2011, simpson2014}. The photometric redshifts estimated by \texttt{HYPERZ} and \texttt{EAZY} agreed within the errors, and no systematic offsets between the two were found. Note that the photometric redshift errors derived by \texttt{HYPERZ} tend to be underestimated because the $\chi^2$ distribution is not a realistic description of the true photometric redshift error distribution \citep{oyaizu2008}. Therefore, we discuss the results from \texttt{EAZY} in the following sections.

Consequently, we obtained the photometric redshifts of SXDF-ALMA1, 2, 4, and 5 ($z_\mathrm{photo} = 2.27^{+0.94}_{-0.87}$, $2.54^{+0.23}_{-0.51}$, $1.33^{+0.10}_{-0.16}$, and $1.52^{+0.13}_{-0.18}$, respectively), while that of SXDF-ALMA3 was poorly constrained ($z_\mathrm{photo} = 2.4^{+2.5}_{-2.0}$) because of the limited number of detections lacking significant break features. The photometric redshifts of SXDF-ALMA1 and 2 were consistent with their narrowband H$\alpha$ redshifts ($z_\mathrm{NB} = 2.53\pm0.02$, see table \ref{table2}) for the \texttt{EAZY} solutions.

\subsubsection{Estimation of stellar masses}
\label{mass}
We estimated the stellar masses of SXDF-ALMA1, 2, 4, and 5, for which reliable SED fits were obtained. The SED fitting at optical-to-near-infrared wavelengths was done by using the \texttt{FAST} code \citep{kriek2009}, which is compatible with \texttt{EAZY}, to derive the stellar masses. In the SED fitting, the templates were taken from the population synthesis model of BC03 with a solar metallicity in accordance with previous research on faint SMGs \citep[e.g.,][]{tadaki2015}. Here, we assumed the Chabrier initial mass function \citep[IMF;][]{chabrier2003}, the Calzetti extinction law, and exponentially declining star formation histories. Following the recipe presented by \citet{wuyts2011}, we used the $e$-folding timescale of SFRs $\tau\geq300$ Myr for fitting. This is because \citet{wuyts2011} suggested that setting $\tau\geq300$ Myr yields the most reasonable SED fits for star-forming galaxies. For SXDF-ALMA1 and 2, we fixed the redshifts to the narrowband redshifts, while we used the photometric redshifts estimated by \texttt{EAZY} for SXDF-ALMA4 and 5. Table \ref{table2} summarizes the results from the SED fitting using \texttt{FAST}.

To constrain the stellar mass of SXDF-ALMA3, we derived it by using a mass-to-light ratio obtained in the rest-frame $H$-band, as done in previous works \citep[e.g.,][]{hainline2011,wardlow2011,simpson2014}. There are several benefits to using a rest-frame $H$-band magnitude. The cooler low-mass stars that dominate the stellar mass of a galaxy emit most of their light at red optical and near-infrared wavelengths. In addition, the rest-frame $H$-band is less sensitive to dust extinction than rest-frame optical bands and is less affected by thermally pulsating asymptotic giant branch stars than the rest-frame $K$-band according to \citet{hainline2011}, who utilized BC03 SED templates with Chabrier IMF and obtained $M_*/L_H$ = 0.17 and 0.13 $\MO\ \LO^{-1}$ for constant and single-burst star formation histories, respectively ($L_H$ is the rest-frame $H$-band luminosity without extinction correction). Here, we adopted the average value $M_*/L_H = 0.15\ \MO\ \LO^{-1}$ of these two extreme cases. If we assumed that SXDF-ALMA3 lies at $z=$ 2, 3, or 4 and used $L_H$ obtained from the rest-frame $H$-band magnitudes in table \ref{table2}, the stellar masses were estimated to be $\sim 5 \times 10^{9}$, $< 2 \times 10^{10}$, and $< 6 \times 10^{10} \MO$, respectively (the inequality sign represents the 3$\sigma$ upper limit). Despite the redshift uncertainty, the constraints favored a lower stellar mass for SXDF-ALMA3 than for the other ALMA sources. This can be one of the reasons why it is faint at optical-to-near-infrared wavelengths.

\subsection{Estimation of SFRs}
\label{sfr}
We computed their SFRs by summing the ultraviolet SFRs (SFR$_\mathrm{UV}$) and infrared SFRs (SFR$_\mathrm{IR}$) based on the work of \citet{kennicutt1998}:
\begin{eqnarray}
\mathrm{SFR_{UV+IR}}(\MO\ \mathrm{yr}^{-1})&=(3.3L_{2800}+L_\mathrm{IR})/\LO\times10^{-10},
\end{eqnarray}
where $L_{2800}$ is the rest-frame 2800 $\mathrm{\AA}$ luminosity. To derive their total infrared luminosities, we used the mid-infrared-to-far-infrared SED templates from \citet{dale2002}, which are often applied to dusty star-forming galaxies observed by {\it Herschel} \citep[e.g.,][]{chapman2010,magnelli2014,ilbert2015}, scaled to the observed flux densities at 1.1 mm ($S_\mathrm{1.1 mm;\ total}$). We assumed the dust temperatures $T_\mathrm{dust}$ = 20--35 K (for SXDF-ALMA4, $T_\mathrm{dust}$ = 20--30 K; see Appendix for details) and an emissivity index $\beta=1.5$, which is a typical value for $z\sim$ 1--2 star-forming galaxies \citep[e.g.,][]{elbaz2011, symeonidis2013}.

To consider a variety of SEDs, we also estimated $L_\mathrm{IR}$ by assuming the template SED of the well-studied starburst galaxy Arp220 \citep{silva1998} and a typical SED of SMGs from \cite{pope2008}. We found that the obtained values were within the uncertainties presented in table \ref{table2}. Note that we only used the observed flux densities at 1.1 mm. Therefore, the derived values should have large systematic uncertainties. We also obtained sSFR by using their ultraviolet + infrared SFRs and stellar masses. Table \ref{table2} summarizes the derived values.

\begin{table*}[]
\caption{Results of the multi-wavelength analysis}
\begin{tabular}{rlccccc}
\hline\hline
&&SXDF-ALMA1&SXDF-ALMA2&SXDF-ALMA3&SXDF-ALMA4&SXDF-ALMA5\\ \hline
$z_{photo}$\footnotemark[$*$]  &(\texttt{EAZY}) &$2.27^{+0.94}_{-0.87}$&$2.54^{+0.23}_{-0.51}$&$2.4^{+2.5}_{-2.0}$&$1.33^{+0.10}_{-0.16}$&$1.52^{+0.13}_{-0.18}$\\
$z_{photo}$\footnotemark[$\dagger$]  &(\texttt{HYPERZ}) &$2.94^{+0.45}_{-0.32}$&$2.06^{+0.21}_{-0.08}$&$3.1^{+3.9}_{-1.8}$&$1.39^{+0.02}_{-0.02}$&$1.63^{+0.03}_{-0.02}$\\
$z_\mathrm{NB}$\footnotemark[$\ddagger$] &&$2.53\pm0.02$&$2.53\pm0.02$&$-$&$-$&$-$\\
$M_*$\footnotemark[$\S$]  &[$\times$ 10$^{10}$ $\MO$]&$9.5^{+2.8}_{-7.3}$&$9.3^{+1.1}_{-6.7}$&$-$&$3.5^{+1.6}_{-0.9}$&$4.1^{+3.2}_{-2.2}$\\
$A_V$\footnotemark[$\S$] &[mag]&$2.6^{+1.9}_{-1.5}$&$1.6^{+0.1}_{-0.8}$&$-$&$1.9^{+0.6}_{-1.0}$&$2.3^{+0.7}_{-1.5}$\\
$L_\mathrm{IR}$\footnotemark[$\|$] &[$\times$ 10$^{12}$ $\LO$]&$2^{+2}_{-1}$&$1^{+1}_{-0.8}$&$1^{+1}_{-0.7}$&$0.3^{+0.2}_{-0.2}$&$0.5^{+0.5}_{-0.4}$\\
SFR$_\mathrm{UV}$\footnotemark[$\#$] &[$\MO$ yr$^{-1}$]&$<2$&$6\pm2$&$-$&$0.6\pm0.2$&$1\pm0.4$\\
SFR$_\mathrm{IR}$\footnotemark[$**$] &[$\MO$ yr$^{-1}$]&$200^{+200}_{-100}$&$100^{+100}_{-80}$&$100^{+100}_{-70}$&$30^{+20}_{-20}$&$50^{+50}_{-40}$\\
SFR$_\mathrm{UV+IR}$\footnotemark[$\dagger\dagger$] &[$\MO$ yr$^{-1}$]&$200^{+200}_{-100}$&$100^{+100}_{-80}$&$-$&$30^{+20}_{-20}$&$50^{+50}_{-40}$\\
sSFR\footnotemark[$\ddagger\ddagger$] &[Gyr$^{-1}$]&$2^{+2}_{-2}$&$1^{+1}_{-1}$&$-$&$0.8^{+0.6}_{-0.6}$&$1^{+2}_{-1}$\\
\hline
\multicolumn{6}{l}{\hbox to 0pt{\parbox{180mm}{{\bf Notes.}{
\\ \footnotemark[$*$] Photometric redshifts with 99\% confidence intervals calculated by using \texttt{EAZY} with flux- and redshift-based priors.
\\ \footnotemark[$\dagger$] Photometric redshifts with 99\% confidence intervals calculated by using \texttt{HYPERZ}.
\\ \footnotemark[$\ddagger$] Narrowband redshifts derived from H$\alpha$ observations \citep{tadaki2013}.
\\ \footnotemark[$\S$] Stellar masses and visual extinction estimated from the SED fitting code \texttt{FAST} with 99\% confidence intervals.
\\ \footnotemark[$\|$] infrared luminosities based on the mid-infrared-to-far-infrared SED templates by \citet{dale2002}. See the Appendix for details.
\\ \footnotemark[$\#$] SFRs obtained from their infrared luminosities.
\\ \footnotemark[$**$] SFRs obtained from their infrared luminosities.
\\ \footnotemark[$\dagger\dagger$] SFR$_\mathrm{UV}$ + SFR$_\mathrm{IR}$
\\ \footnotemark[$\ddagger\ddagger$] sSFRs obtained from their SFR$_\mathrm{UV+IR}$ and stellar masses.
}}
\hss}}
\label{table2}
\end{tabular}
\end{table*}

\subsection{Optical-to-radio SED}
\label{m-r_sed}
\begin{figure*}[!ht]
\begin{center}
\FigureFile(160mm,50mm){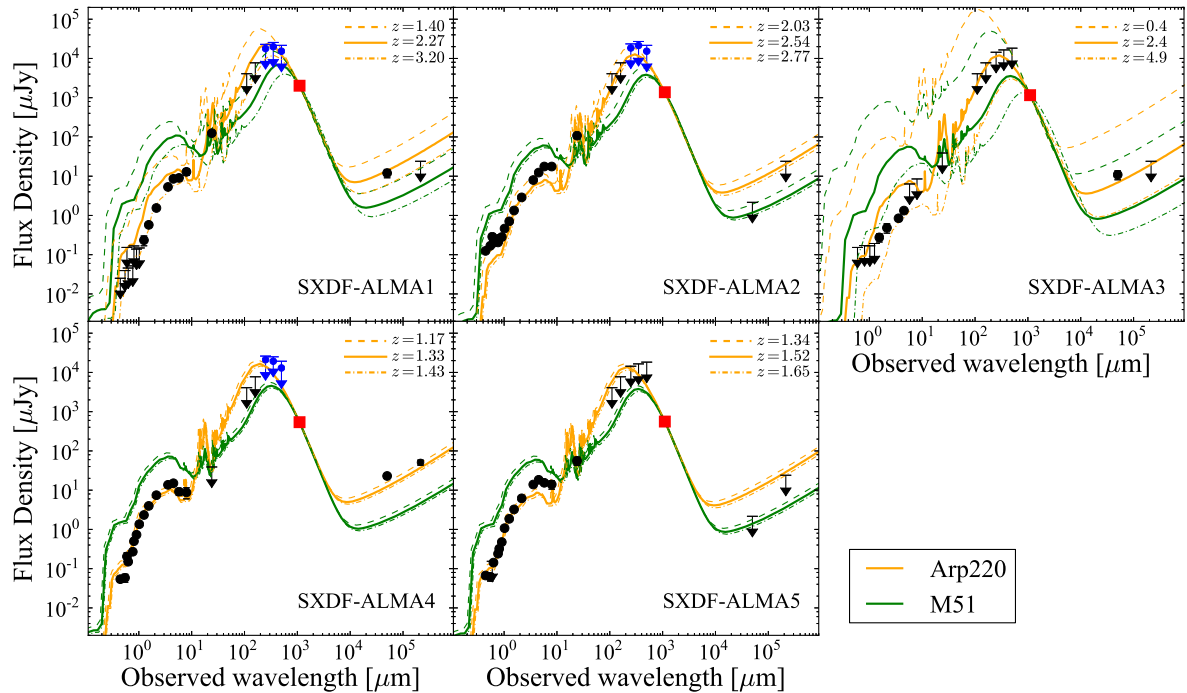}
\end{center}
\caption{Optical-to-radio SED of SXDF-ALMA1, 2, 3, 4, and 5. The red squares show 1.1-mm flux densities. The black dots indicate photometric data points at the optical-to-radio wavelengths. The black arrows represent 3$\sigma$ upper limits. When SPIRE photometry suffered from heavy contamination of nearby sources (SXDF-ALMA1, 2, and 4), we plotted nominal values of SPIRE photometry as upper limits (see \S 3.2). For comparison, we plotted SED templates of Arp220 and M51 \citep{silva1998}. The SED templates were scaled to the flux densities at 1.1 mm. The redshifts were fixed to the best-fit values estimated by \texttt{EAZY} (solid lines) and the lower and upper limits of the 99\% confidence intervals. }
\label{full_sed}
\end{figure*}

Figure \ref{full_sed} plots the optical-to-radio SEDs of the ALMA sources.
Because SXDF-ALMA1, 2, and 4 suffered from heavy contamination in the Herschel SPIRE photometry (see section \ref{photo_mir-r} and figure \ref{multi_ww}), we plotted nominal values of SPIRE photometry with the blue dots with arrows, which should be regarded as upper limits. For comparison, we also plotted two different template SEDs modeled by \citet{silva1998}: a dusty starburst galaxy (Arp220) and spiral galaxy (M51). They also differ in dust temperature: $T_{\rm dust}=47$ K for Arp220 \citep{Klaas1997} and $T_{\rm dust}=24.9$ K for M51 \citep{mentuch2012}. For the solid lines, the redshifts were fixed to the best-fit photometric redshifts estimated by \texttt{EAZY}. To account for the redshift uncertainties, we also plotted the template SEDs for which the redshift was fixed to the lower and upper boundaries of the 99\% confidence intervals. These SEDs were scaled to the flux densities at 1.1 mm.

As shown in figure \ref{full_sed}, the photometric redshifts from optical-to-near-infrared photometry and resulting SEDs extended to far-infrared produced a reasonable model for the dust emission. The optical-to-radio SEDs of the ALMA sources seems to match Arp220 better than M51. In particular, the rest-frame ultraviolet+optical SEDs show evidence for more obscured young stellar population than the local disk galaxies like M51, as expected. Previously, \cite{dacunha2015} have suggested that the average SED of SMGs identified in the ALMA follow-up observation of the LABOCA Extended Chandra Deep Field South surveys \citep[ALESS;][]{hodge2013} is inconsistent with Arp220, especially optical to near-infrared wavelengths. The SEDs of these ALMA sources are more like the local ultra-luminous infrared galaxies than ALESS SMGs. 

As shown in figure \ref{multi_ww}, SXDF-ALMA1, 2, and 4 were heavily blended with nearby MIPS sources in the {\it Herschel}/SPIRE bands, which suggests that their flux densities were overestimated. At $z\sim$ 1.3, a silicate absorption feature at rest frame 9.7 $\micron$ shifted into the MIPS 24 $\micron$ band. This may be why SXDF-ALMA4 was not detected at MIPS 24 $\micron$. The local (ultra-)luminous infrared galaxies exhibited a broad silicate absorption feature at rest frame 9.7 $\micron$ \citep[e.g.,][]{armus2007, pereira2010}. The silicate absorption feature is known to merely require a mass of warm dust obscured by a significant column of cooler dust \citep[e.g.,][]{magdis2011}. SXDF-ALMA4 was detected at 1.4 GHz, and the photometry is consistent with the SED template of Arp220. This suggests that the dust temperature is higher than the assumed $T_\mathrm{dust}$ = 20--30 K (see the Appendix for details).

The far-infrared-to-radio SED of SXDF-ALMA3 may place a more stringent constraint on the redshift than the optical/near-infrared photometric one. As shown in figure \ref{full_sed}, the upper limits and 6 GHz photometry suggest that this object is located at $z\simeq$ 2--3 if the SED is similar to M51 and Arp220.


\begin{figure*}[!ht]
\begin{center}
\FigureFile(160mm,50mm){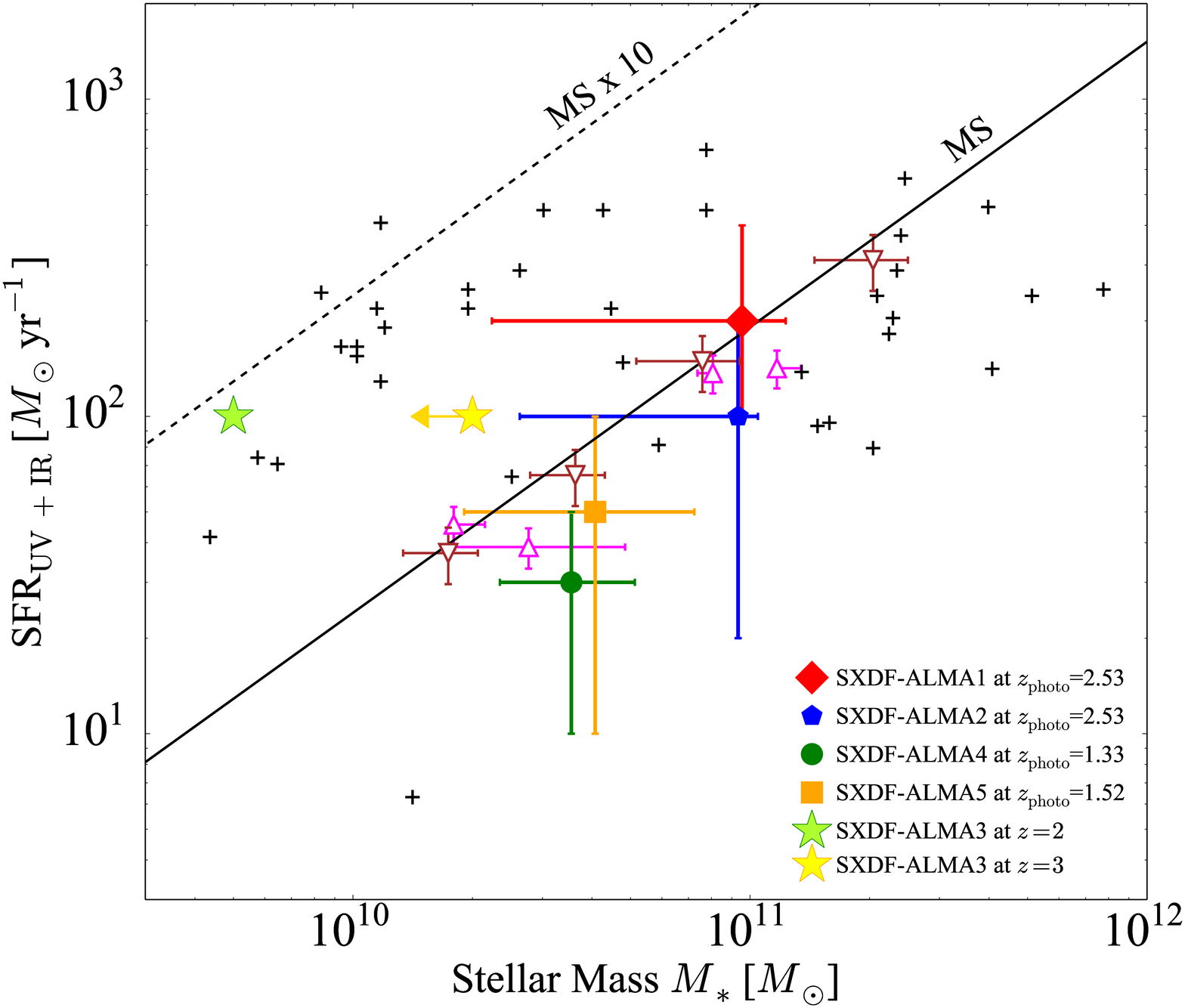}
\end{center}
\caption{SFRs derived from ultraviolet and infrared luminosities plotted against their stellar masses. The diamond, hexagon, circle, and square symbols represent SXDF-ALMA1, 2, 4, and 5, respectively. The solid line indicates the main sequence for star-forming galaxies at $1.4<z<2.5$ as defined by \citet{daddi2007}. The dashed line indicates the loci 10 times above the main sequence. The black crosses indicate ALESS sources at $1.3 < z < 2.5$ from \citet{dacunha2015}. The magenta triangles are faint 1.3 mm sources detected by ALMA \citep{hatsukade2015}. Note that \citet{hatsukade2015} used 68\% confidence intervals, whereas we used 99\% confidence intervals. The brown inverted triangles are the average values of BzK galaxies derived from the PACS stacking analysis \citep{rodighiero2015}. The green and yellow stars indicate when SXDF-ALMA3 lay at $z=2$ and SXDF-ALMA3 lay at $z=3$, respectively. Note that the stellar mass of SXDF-ALMA3 at $z=3$ was at the 3$\sigma$ upper limit because the rest-frame $H$-band was not detected.
}
\label{m-sfr}
\end{figure*}

\section{Discussion}
\label{discussion}
\subsection{Contribution to the infrared extragalactic background light}
Owing to the high sensitivity and high angular resolution observations with ALMA, 50\%--100\% of the infrared extragalactic background light has been claimed to be resolved if we go down to $\sim0.1$--0.02 mJy \citep[e.g.,][]{hatsukade2013, ono2014, carniani2015, fujimoto2015}. From the summation of the 1.1 mm flux densities of all of our ALMA sources and the survey area (2 arcmin$^2$), the contribution of the ALMA sources to the infrared extragalactic background light was estimated to be$\sim$ $10^{+6}_{-4}$ Jy deg$^{-2}$, which corresponds to $\sim 40^{+24}_{-16}\%$ of the infrared extragalactic background light obtained by \citet{fixsen1998} using the COsmic Background Explore (COBE) satellite \citep[25$^{+22}_{-13}$ Jy deg$^{-2}$;][]{carniani2015} or $\sim54^{+32}_{-22}\%$ if we adopt the the COBE measurement made by \citet[][18.5 Jy deg$^{-2}$]{puget1996}. In the subsequent 
discussion, we adopt the Fixsen et al. value for the infrared extragalactic background light, but we caution that there exist uncertainties (likely systematic) in the COBE measurements. Because of the numbers of our sources, we used the Poisson uncertainty values presented by \citet{gehrels1986}. The completeness in the flux range of ALMA sources was $\sim100\%$ \citep{hatsukade2016}.

Note that this value can be overestimated because our observation field was selected to include a single bright SMG (Ikarashi et al.~in preparation) and a chain of HAEs at $z=2.5$ \citep{tadaki2013,tadaki2015}. Given that SXDF-ALMA1 and 2 were identified as an AzTEC/ASTE source (Ikarashi et al.~in preparation) and HAEs \citep{tadaki2013}, it is better to exclude SXDF-ALMA1 and 2 when discussing the real contribution of the ALMA sources to the infrared extragalactic background light. The contributions of SXDF-ALMA3, 4, and 5 to the infrared extragalactic background light were estimated to be $\sim$ $4.1^{+5.4}_{-3.0}$ Jy deg$^{-2}$, which corresponds to $\sim 16^{+22}_{-12}\%$ of the infrared extragalactic background light obtained by the COBE satellite. This suggests that their contribution to the infrared extragalactic background light is as large as that of bright SMGs \citep[$S_\mathrm{1.1mm}\geq$ 1.0 mJy, $\sim$ 2.9 Jy deg$^{-2}$;][]{hatsukade2011}. Although our survey area was small and may have been affected by cosmic variance, these results suggest that bright ($S_\mathrm{1.1mm}\geq$ 1 mJy) sources and faint SMGs with 0.5 mJy $< S_\mathrm{1.1 mm}\lesssim$ 1.0 mJy, which is the flux range of the ALMA source, seem to contribute $\sim 28^{+22}_{-12}\%$ to the infrared extragalactic background light. These results suggest that faint SMGs with $S_\mathrm{1.1 mm}\lesssim$ 0.5 mJy are major contributors to the infrared extragalactic background light. The results of stacking analysis of near-infrared selected galaxies with $S_\mathrm{1.1 mm}\lesssim$ 0.5 mJy will be discussed in our upcoming paper (Wang et al.~in preparation).

\subsection{Contribution to the cosmic SFRD}
Substantial attempts have been made with {\it Herschel} to resolve the redshift evolution of the contribution of infrared selected galaxies up to the $z=$ 3 \citep{brugarella2013}. \citet{brugarella2013} estimated the cosmic infrared SFRD from the infrared luminosity functions inferred from {\it Herschel} observations and found that the contribution peaks at $z=1.35$, which accounts for 89\% of the total cosmic SFRD of $1.1\times10^{-1}$ $\MO$ yr$^{-1}$ Mpc$^{-3}$ (using Chabrier IMF). However, they used the extrapolate infrared luminosity functions below the confusion limit of {\it Herschel} (e.g., $L_\mathrm{IR}\lesssim10^{12}\ \LO$ at $z\simeq 2$) to estimate the cosmic infrared SFRD. \citet{wardlow2011} derived infrared luminosity functions of SMGs detected by the LABOCA Extended Chandra Deep Field South surveys \citep[LESS;][]{weiss2009}. However, they also did not investigate the luminosity range of $L_\mathrm{IR}\lesssim10^{12}\ \LO$.

ALMA sources with (photometric) redshifts allow the constraint on the contribution from a faint ($L_\mathrm{IR}\lesssim10^{12}\LO$) population of star-forming galaxies. In addition, the contribution from galaxies undetected by {\it Herschel} can be estimated. We simply assume that the ALMA sources lie in the redshift interval of $1<z<4$ (co-moving volume: $V_{\rm com}\sim$ 1.9 $\times$ 10$^{4}$ Mpc$^3$) to cover the redshift uncertainties of the ALMA sources. Then, we estimate the contribution from all of the ALMA sources, including SXDF-ALMA1 and 2, on the basis of SFRs simply derived from the 1.1 mm flux densities (section \ref{sfr}). Considering the uncertainty in $T_\mathrm{dust}$ (see section \ref{sfr} and Appendix), the inferred infrared SFRD for $1 < z < 4$ is $\sim$ (0.9--5) $\times10^{-2}$ $\MO$ yr$^{-1}$ Mpc$^{-3}$, which accounts for $\sim$ 10--70\% of the average infrared SFRD at $0.9<z<3.6$ as estimated by {\it Herschel} \citep[using Chabrier IMF, $\sim 7\times10^{-2}$ $\MO$ yr$^{-1}$ Mpc$^{-3}$;][]{brugarella2013}. If we exclude SXDF-ALMA1 and 2 to avoid the contribution of the known AzTEC source, then the infrared SFRD for $1 < z < 4$ is estimated to be $\sim$ (0.3--2) $\times10^{-2}$ $\MO$ yr$^{-1}$ Mpc$^{-3}$. The inferred cosmic infrared SFRD is similar to that of bright SMGs ($S_\mathrm{870\micron}>$ 4 mJy) at $z\simeq$ 2--3 [$\sim$ (1--2) $\times10^{-2}$ $\MO$ yr$^{-1}$ Mpc$^{-3}$ using Chabrier IMF; \cite{wardlow2011}]. These results imply that the ALMA sources play an important role in the cosmic SFRD, even if we exclude the contribution of the known AzTEC source at $1 < z < 4$. Note that our results can be affected by the cosmic variance and clustering because of our small survey area. Therefore, future ALMA large surveys will provide a stronger constraint on the role of faint SMGs in the cosmic SFRD.

\citet{chen2014} inferred that there are many submillimeter sources that are difficult to detect in deep optical/near-infrared surveys like SXDF-ALMA3. Indeed, submillimeter sources which have no counterparts at optical/near-infrared wavelengths have been reported \citep[e.g.,][]{wang2007, smolcic2012, simpson2014, dunlop2016}.
However, their real contributions to the cosmic infrared SFRD is still uncertain. The contribution of SXDF-ALMA3 to the cosmic infrared SFRD may be $\sim$ 0.1--1 $\times10^{-2}$ $\MO$ yr$^{-1}$ Mpc$^{-3}$ or $\sim$ 1\%--10\% of the average infrared SFRD at $0.9<z<3.6$ as estimated by {\it Herschel} ($\sim 7\times10^{-2}$ $\MO$ yr$^{-1}$ Mpc$^{-3}$) if this object lies somewhere in the redshift interval of $1<z<4$.


\subsection{Star formation properties of the ALMA detected sources}
Determining the star-forming properties of faint SMGs is important to understand the evolution of the cosmic SFRD because they are main contributors to the cosmic SFRD. We investigated the star formation mode of faint SMG counterparts to check whether or not their star-forming properties are similar to starburst galaxies or not. 

Figure \ref{m-sfr} plots the total SFRs (SFR$_\mathrm{UV}$ + SFR$_\mathrm{IR}$) of SXDF-ALMA1, 2, 4, and 5 as functions of their stellar mass.
We also show the average values of {\it BzK} galaxies derived by the PACS stacking analysis \citep{rodighiero2015}, SMGs identified in ALESS surveys \citep[][]{hodge2013} at $1.3 < z < 2.5$, and faint 1.3 mm sources detected with ALMA \citep{hatsukade2015}. We plotted SFRs of the ALESS sources obtained by \citet{dacunha2015} by fitting SEDs at ultraviolet to radio wavelengths. This figure shows that SXDF-ALMA1, 2, 4, and 5 are located in the main sequence. This means that they are more like ``normal'' star-forming galaxies rather than extremely starburst galaxies. These results are consistent with those of \citet{koprowski2014} and \citet{hatsukade2015}. Note that the total SFR of SXDF-ALMA1 should be treated as an upper limit because we used the 3$\sigma$ upper limit value as its SFR$_\mathrm{UV}$. However, this does not affect our results because the SFR$_\mathrm{UV}$ of SXDF-ALMA1 is negligible compared to its SFR$_\mathrm{IR}$ (see table \ref{table2}).

Figure \ref{m-sfr} also shows the constraints of the stellar mass and SFR of SXDF-ALMA3. The results imply that SXDF-ALMA3 is an starburst galaxy with a small stellar mass compared to bright SMGs \citep[$M_*\sim9.0\times10^{10}\ \MO$ using Chabrier IMF;][]{hainline2011}. Submillimeter sources such as SXDF-ALMA3 have been missed in previous deep optical/NIR surveys and submillimeter single-dish surveys. Future spectroscopic identification of such sources using ALMA is highly encouraged.

Finally, we compared our results with the theoretical predictions obtained by recent simulations and semi-analytical models. \citet{bethermin2012} empirically predicted the number counts at far-infrared and millimeter wavelengths from mid-infrared and radio number counts and suggested that galaxies with $S_\mathrm{1.1 mm}\lesssim$ 1 mJy are more likely to be associated with main sequence star-forming galaxies by using the SED library based on {\it Herschel} observations. From a theoretical point of view, \citet{hayward2013} predicted the number counts at submillimeter wavelengths from a semi-empirical model with three-dimensional hydrodynamical simulations and three-dimensional dust radiative transfer and also suggested that galaxies with $S_\mathrm{1.1 mm}<$ 1 mJy are more likely to be associated with main sequence star-forming galaxies. These predictions are consistent with our results that two of the three faint SMGs (SXDF-ALMA4 and 5) are main sequence star-forming galaxies, as shown in figure \ref{m-sfr}.

\section{Summary}
\label{summary}
We detected five submillimeter sources ($S_\mathrm{1.1mm}$ = 0.54--2.02 mJy) in the SXDF-UDS-CANDELS field by using ALMA. The two brightest sources correspond to a known single-dish (AzTEC) selected bright SMG, whereas the remaining three are faint SMGs newly uncovered by ALMA. Our main results are as follows:
\begin{itemize}
\item If we exclude SXDF-ALMA1 and 2 to avoid the contribution of the known AzTEC source, the contribution of the faint SMGs to the infrared extragalactic background light is estimated to be $\sim$ $4.1^{+5.4}_{-3.0}$ Jy deg$^{-2}$, which corresponds to $\sim$ $16^{+22}_{-12}$\% of the infrared extragalactic background light. This suggests that their contribution to the infrared extragalactic background light is as large as that of bright SMGs \citep[$S_\mathrm{1.1mm}\geq$ 1.0 mJy, $\sim$ 2.9 Jy deg$^{-2}$;][]{hatsukade2011}. 

\item The infrared SFRD of SXDF-ALMA3, 4, and 5 for $1 < z < 4$ is estimated to be $\sim$ (0.3--2) $\times10^{-2}$ $\MO$ yr$^{-1}$ Mpc$^{-3}$. This value is as large as the contribution to infrared SFRD of bright SMGs at $z\simeq$ 2--3.

\item For four of the five ALMA sources (SXDF-ALMA1, 2, 4, and 5), we obtained the photometric redshifts and $M_*$ by SED fitting at optical-to-near-infrared wavelengths: $z_\mathrm{photo}$ $\simeq$ 1.3--2.5, $M_*$ $\simeq$ (3.5--9.5) $\times10^{10}\ \MO$. The SFRs were estimated from ultraviolet and infrared luminosities as follows: SFR$_\mathrm{UV+IR}\simeq$ 30--200 $\MO$ yr$^{-1}$. We also obtained their sSFRs ($\simeq$ 0.8--1 Gyr$^{-1}$). The derived values indicate that they are more like ``normal'' star-forming galaxies than starburst galaxies.

\item SXDF-ALMA3 is faint in the optical-to-near-infrared (\textit{H} = 25.30 $\pm$ 0.25), despite its infrared luminosity ($L_\mathrm{IR}\simeq 1\times10^{12} \LO$ or SFR $\simeq$ 100 $\MO$ yr$^{-1}$). The optical-to-radio SED suggests that this object is located at $z\simeq$ 2--3. The inferred stellar mass ($M_*$ $\sim$ 5 $\times$ 10$^{9}$ or $<$ 2 $\times$ 10$^{10}$ at $z=2$ or $z=3$, respectively) was likely to be smaller than that of the other ALMA sources. These results suggest that this object may be an starburst galaxy with a small stellar mass at $z\simeq$ 2--3. The contribution of SXDF-ALMA3 to the cosmic SFRD may be $\sim$ 1\%--10\% of the infrared SFRD.
\end{itemize}

\bigskip

We thank the referee for the comments, which improve the manuscript.
This paper makes use of the following ALMA data: ADS/JAO.ALMA\#2012.1.00756.S.
ALMA is a partnership of ESO (representing its member states), NSF (USA), and NINS (Japan) together with NRC (Canada), NSC and ASIAA (Taiwan), and KASI (Republic of Korea) in cooperation with the Republic of Chile.
The Joint ALMA Observatory is operated by ESO, AUI/NRAO, and NAOJ.
This research is supported by the ALMA Japan Research Grant of NAOJ Chile Observatory, NAOJ-ALMA-0049 and NAOJ-ALMA-0099.
Data analysis was partly carried out on the common-use data analysis computer system at the Astronomy Data Center (ADC) of the National Astronomical Observatory of Japan.
Y.~Yamaguchi and H.~Umehata are thankful for the JSPS fellowship.
K.~Kohno, Y.~Matsuda, B.~Hatsukade, and R.~Makiya acknowledge support from JSPS KAKENHI Grant Numbers 25247019, 20647268, 15K17616 and 15H05896, respectively.
J.~S.~Dunlop acknowledges the support of the European Research Council via an Advanced Grant.
%
%
%
I.~Aretxaga is supported by the CONACYT grant number CB-2011-01-167291.
This research made use of the NASA/IPAC Extragalactic Database (NED), which is operated by the Jet Propulsion Laboratory, California Institute of Technology, under contract with the National Aeronautics and Space Administration.

\begin{appendix}
\section*{Estimation of infrared luminosities}
\begin{figure*}[t]
\begin{center}
\FigureFile(160mm,50mm){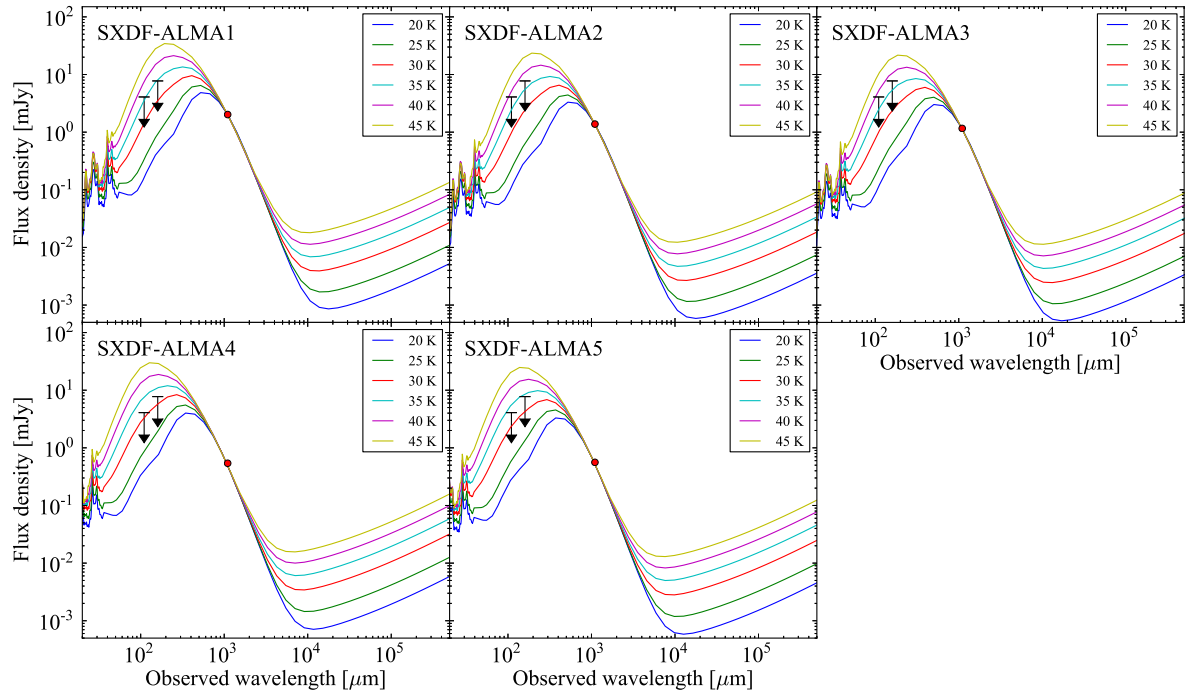}
\end{center}
\caption{SED templates by \citet{dale2002} scaled to the observed flux densities at 1.1 mm ($S_\mathrm{1.1mm\ total}$). The blue, green, red, cyan, magenta, and yellow solid lines indicate SED templates with $T_\mathrm{dust}$ = 20, 25, 30, 35, 40, and 45 K, respectively. The red circles are the observed flux densities at 1.1 mm. The black arrows represent the 3$\sigma$ upper limits on {\it Herschel}/PACS 100 $\micron$ and 160 $\micron$.
}
\label{dale_temp}
\end{figure*}

As described in section \ref{sfr}, we used the SED templates obtained by \citet{dale2002} to estimate infrared luminosities of the ALMA sources. Figure \ref{dale_temp} plots the SED templates scaled to the observed flux densities at 1.1 mm ($S_\mathrm{1.1mm\ total}$). Here, we show the SED templates with $T_\mathrm{dust}$ = 20, 25, 30, 35, 40, and 45 K. The ALMA sources were not detected at {\it Herschel}/PACS 100 $\micron$ and 160 $\micron$. As shown in the figure, the 3$\sigma$ upper limits on {\it Herschel}/PACS photometry could place a stringent constraint on $T_\mathrm{dust}$. The upper limits on the {\it Herschel}/PACS photometry are clearly below the SED templates with $T_\mathrm{dust}$ = 40 and 45 K. For SXDF-ALMA4, they are clearly below the SED templates with $T_\mathrm{dust}$ = 35, 40, and 45 K. Therefore, we assumed $T_\mathrm{dust}$ = 20--35 K ($T_\mathrm{dust}$ = 20--30 K for SXDF-ALMA4) to estimate the infrared luminosities of ALMA sources.

For each ALMA source, we integrated SED templates with $T_\mathrm{dust}$ = 20, 25, 30, and 35 K ($T_\mathrm{dust}$ = 20, 25, and 30 K for SXDF-ALMA4) between the rest-frames 8 and 1000 $\micron$ and adopted the average, minimum, and maximum values as the infrared luminosity, lower limit of the infrared luminosity, and upper limit of the infrared luminosity, respectively.

\end{appendix}

%

\end{document}